\documentclass[aps,prd,a4paper,reprint,amsmath,amssymb,superscriptaddress,nofootinbib]{revtex4-2}
\usepackage[utf8]{inputenc}
\usepackage{graphicx} % Required for inserting images
\usepackage{calligra,amsmath,amsfonts,bbm,mathrsfs,amssymb}
\usepackage{graphicx}
\usepackage{caption}
\usepackage{enumerate}
\usepackage{hyperref}
\usepackage{braket}
\usepackage{physics}
\usepackage[dvipsnames]{xcolor}
\usepackage{colortbl}
\usepackage{tabularx}

\newcommand{\X}{\mathcal{X}}

\begin{document}

\title{Axion Detection Experiments Meet the Majoron}

\author{Qiuyue Liang}\email{qiuyue.liang@ipmu.jp}
\affiliation{Kavli IPMU (WPI), UTIAS, University of Tokyo, Kashiwa, 277-8583, Japan}

\author{Xavier Ponce Díaz}\email{xavier.poncediaz@pd.infn.it}
\affiliation{Istituto Nazionale di Fisica Nucleare (INFN), Sezione di Padova, and Dipartimento di Fisica e Astronomia ``G.~Galilei'', Universit\`a di Padova, Via F. Marzolo 8, 35131 Padova, Italy}

\author{Tsutomu T. Yanagida}\email{tsutomu.tyanagida@sjtu.edu.cn}
\affiliation{Kavli IPMU (WPI), UTIAS, University of Tokyo, Kashiwa, 277-8583, Japan}
\affiliation{Tsung-Dao Lee Institute \& School of Physics and Astronomy, Shanghai Jiao Tong University, China}

\begin{abstract}
    The majoron is a well-motivated light (pseudo-Nambu-Goldstone) boson associated with the spontaneous breaking of a global lepton-number symmetry. In this {\it letter}, we relate the spontaneous breaking scale and its soft-breaking mass by requiring that the majoron is the main component of the dark matter. An electromagnetic-anomalous coupling can be induced by minimally modifying the original majoron model, surprisingly, predicting a parameter region that largely overlaps with the QCD-axion dark matter band. Thus, we expect that axion search experiments meet the majoron.   
 \end{abstract}

\maketitle

\section{Introduction}
The heavy Majorana (right-handed) neutrinos are key ingredients for generating small neutrino masses via the seesaw mechanism \cite{Minkowski:1977sc,Yanagida:1979as, Yanagida:1979gs,Glashow:1979nm,Gell-Mann:1979vob} \footnote{After the discovery of the seesaw mechanism many intense discussions on the origin of neutrino masses followed \cite{Weinberg:1979sa,  Wilczeck:1979CP, Barbieri:1979ag,Witten:1979nr, Mohapatra:1979ia}.} 
and for producing the baryon-number asymmetry in the universe through the leptogenesis \cite{Fukugita:1986hr,Buchmuller:2005eh}. However, the origin of the heavy Majorana masses remains unknown. The most popular mechanism to generate heavy Majorana mass terms of the right-handed neutrinos is given by spontaneous breaking of the $B-L$ gauge symmetry\cite{Wilczeck:1979CP,Barbieri:1979ag}, where the presence of the three right-handed neutrinos is motivated to cancel the possible gauge anomalies.

There is, however, an alternative theory where the spontaneous breaking of a lepton number $U(1)_L$ \cite{Chikashige:1980qk,Chikashige:1980ui} induces the heavy Majorana masses\footnote{See \cite{Gelmini:1980re} for a low scale lepton-number breaking.}. This symmetry must be a global symmetry, since $U(1)_L$ is anomalous, therefore, cannot be gauged. Although this model is less attractive, there is a big advantage: we have a light pseudo Nambu-Goldstone boson called ``majoron $J$'' which can be an interesting candidate for Dark Matter (DM) \cite{Gu:2010ys,Frigerio:2011in}.

% In this letter, we will study this possibility. 

It is believed that global symmetries cannot be exact, and the majoron could gain a mass by introducing a soft symmetry breaking term \cite{Giddings:1987cg,Alonso:2017avz,Coleman:1985rnk}. In this {\it letter}, to address the dark matter problem, we will consider the majoron mass fixed by the misalignment mechanism \cite{Abbott:1982af,Dine:1981rt,Preskill:1982cy} to get the right DM relic abundance, but not specify the UV origin of this mass. 

Existing literature has explored the DM majoron mass primarily above MeV region \cite{Garcia-Cely:2017oco,Akita:2023qiz,deGiorgi:2023tvn}, and at keV region\cite{Heeck:2017xbu}. This focus is driven by the possibility of searching for such a particle: majorons heavier than the MeV could be detected in neutrino experiments \cite{Garcia-Cely:2017oco,Akita:2023qiz}, whereas below this threshold the detection becomes challenging, see \cite{Bloch:2020uzh} for discussion in keV band. However, to serve the purpose of seesaw mechanism and leptogenesis, the decay constant of the majoron model is given by, $F_J\sim 10^{13} \,\textrm{GeV}$ \cite{Buchmuller:2005eh}. On top of this, to explain the dark matter abundance, the desired majoron mass is, $m_J\sim  \mu$eV. This significantly deviates from the typical majoron detectable region, and therefore, should be searched by different experiments. In this {\it letter},  we introduce a second Higgs boson to enable opposite $U(1)_L$ charges to the leptons, making the model anomalous under QED, and therefore, can be detected by various axion experiments. Moreover, the majoron preferred region overlaps with the QCD-axion dark matter band, and will meet in the way of future axion detection experiments.  

In the rest of this paper, we will introduce the relevant part of the model in Section~\ref{sec:MajoronModel}, and discuss the majoron dark matter in Section~\ref{sec:MajoronDM}. The conclusions and discussions are given in Section~\ref{sec:conclusions}, with an overlook on the possible connections of this model with other phenomena. Finally, this letter is accompanied by an Appendix~\ref{sec:appendix} with some details on the computations.

\section{Majoron model}
\label{sec:MajoronModel}
In this section we briefly describe our majoron model based on a modified lepton-number $U(1)_L$ symmetry.

The key aspect is that the global $U(1)_L$ symmetry-breaking induces the large Majorana masses for the right-handed neutrinos $N_R$. Thus, we consider the right-handed neutrinos $N_R$ carry the $U(1)_L$ charge, fixing it to be $\X_N= +1/2$.  Accordingly, we introduce a complex scalar
$\phi $, whose vacuum-expectation value (vev) $\langle \phi \rangle =v_\phi/\sqrt{2}$ breaks the global $U(1)_L$ symmetry inducing the Majorana masses, and the phase of the scalar field becomes a massless pseudo Nambu-Goldstone boson, which we refer to as {\it majoron} in the remaining context. We assign the charge of $\phi$ to be $\X_\phi= 1$ so that we have a coupling
\begin{equation}
   \mathcal{L}\supset \bar{\ell}_L Y_e e_R H+\bar{\ell}_L Y_D N_R \widetilde{H}+ \frac{1}{2}\bar{N}^c_RY_N   N_R \phi^* \ ,
\end{equation}
where $H$ denotes the Higgs field, and $Y_i, i\in \{e,D,N\}$ are the Yukawa coupling constants. Here, we see the Majorana mass for the right-handed neutrinos is $M_N=Y_N v_\phi/\sqrt{2}$. For typical right-handed neutrino mass that can gives rise to correct neutrino mass in standard model, $M_N \sim 10^{14}$GeV, the favoured vev is set to be the same magnitude assuming a $\mathcal{O}(1)$ coupling constant. The above coupling also fixed the lepton charge of right-handed electron $e_R$ to be $\X_e =1/2$ \footnote{Notice we assign the charge of the right-handed electrons to be the same as the right-handed neutrinos. This is a reason we call the global symmetry as the lepton-number symmetry $U(1)_L$.}, and that of the left-handed lepton doublets $\ell_L$ to be $\X_\ell =1/2$. This is  the common lepton number assignment for majoron models \cite{Chikashige:1980qk}. However, one can check that this model does not has QED anomaly, and therefore cannot be detected through axion experiments.

In this {\it letter}, we introduce the second Higgs field, $H_2$, to assign different lepton charges to $\ell_L$ and $e_R$ to obtain the QED anomaly. The relevant part of the Lagrangian in the lepton sector is given by
\begin{align}
\label{eq:Lagrangian1}
    \mathcal{L}= &\bar{\ell}_L Y_e e_R H_2  +\bar{\ell}_L Y_D N_R \widetilde{H}_1 + \frac{1}{2}\bar{N}_R^c Y_N N_R \phi^* \nonumber\\&+\mu_\phi H_2^\dagger H_1 \phi +V(H_1,\,H_2,\, \phi)+ \text{h.c.}\, , 
\end{align}
where the two Higgs fields carry different lepton charges, $\X_2 -\X_1 = \X_\phi =1$. In this model, the lepton charge of the right-handed electrons is fixed to be $\X_e = -1/2$, while that of the lepton doublet is chosen to be $\X_\ell =1/2$. One can then compute the anomaly coefficients as 
\begin{eqnarray}
\label{eq:Anomaly}
    \mathcal{L}_{\textrm{anom.}} &=& n_f (\X_\ell-\X_e)\frac{\alpha_{\textrm{em}}}{4 \pi} \frac{J}{F_J} F_{\mu\nu}\widetilde{F}^{\mu\nu} \nonumber\\
    &=&   3\frac{\alpha_{\textrm{em}}}{4 \pi} \frac{J}{F_J} F_{\mu\nu}\widetilde{F}^{\mu\nu}  = \frac{g_{J\gamma\gamma}}{4} J F_{\mu\nu}\widetilde{F}^{\mu\nu}  \ ,    
\end{eqnarray}
where $n_f$ denotes the lepton family number, and $F_J$ is the decay constant of majoron model. To obtain the normalised kinetic term for majoron, $F_J$ is set to be the same as the scalar vev, $v_\phi$. This anomalous coupling can be detected through axion experiments that we will further discuss in the next section. It is worth pointing out that this coupling constant, $g_{J\gamma\gamma}$, coincides with the prediction of cosmological birefringence, see \cite{Lin:2022niw}.

Finally, before treating the majoron DM in the next section, we want to briefly comment on the model Eq.\eqref{eq:Lagrangian1}. Although for simplicity we chose the charges such that the quark sector remains uncharged, they are generally charged and can be included in the Lagrangian as discussed in the Appendix~\ref{sec:appendix}. We also note that in the literature several variations of this model exist, for instance, where the majoron becomes an axion, solving the strong-CP problem \cite{Langacker:1986rj, Clarke:2015bea,Sopov:2022bog}, as an ALP \cite{DiLuzio:2024jip} without including the RH neutrinos, or in a 2 Higgs Doublet model with no scalar field $\phi$ \cite{Clarke:2015hta}. However, to the best of our knowledge, this majoron variation had not been presented yet.
 
\section{Majoron dark matter}
\label{sec:MajoronDM}
One of the most pressing problems of particle cosmology, if not the most, is the explanation of the Dark Matter component of the universe. In this respect, light particles have a very appealing mechanism to saturate the relic abundance of DM, firstly proposed for the axion, known as the misalignment mechanism \cite{Dine:1981rt,Abbott:1982af,Preskill:1982cy}. Since we do not specify the UV origin of the majoron mass, we apply the misalignment mechanism directly and estimate the fractional energy density of DM as \cite{Blinov:2019rhb}, 
\begin{align}
\label{eq:DM}
    \Omega_J h^2 
    &\simeq 0.12 \left(\frac{m_J}{\mu\textrm{eV}}\right)^{1/2} \left(\frac{F_J  \theta^i_J   }{1.9\times  10^{13}\,\textrm{GeV}}\right)^2\ ,
\end{align}
\begin{figure*}[tb]
    \centering
    \includegraphics[width=0.85\textwidth]{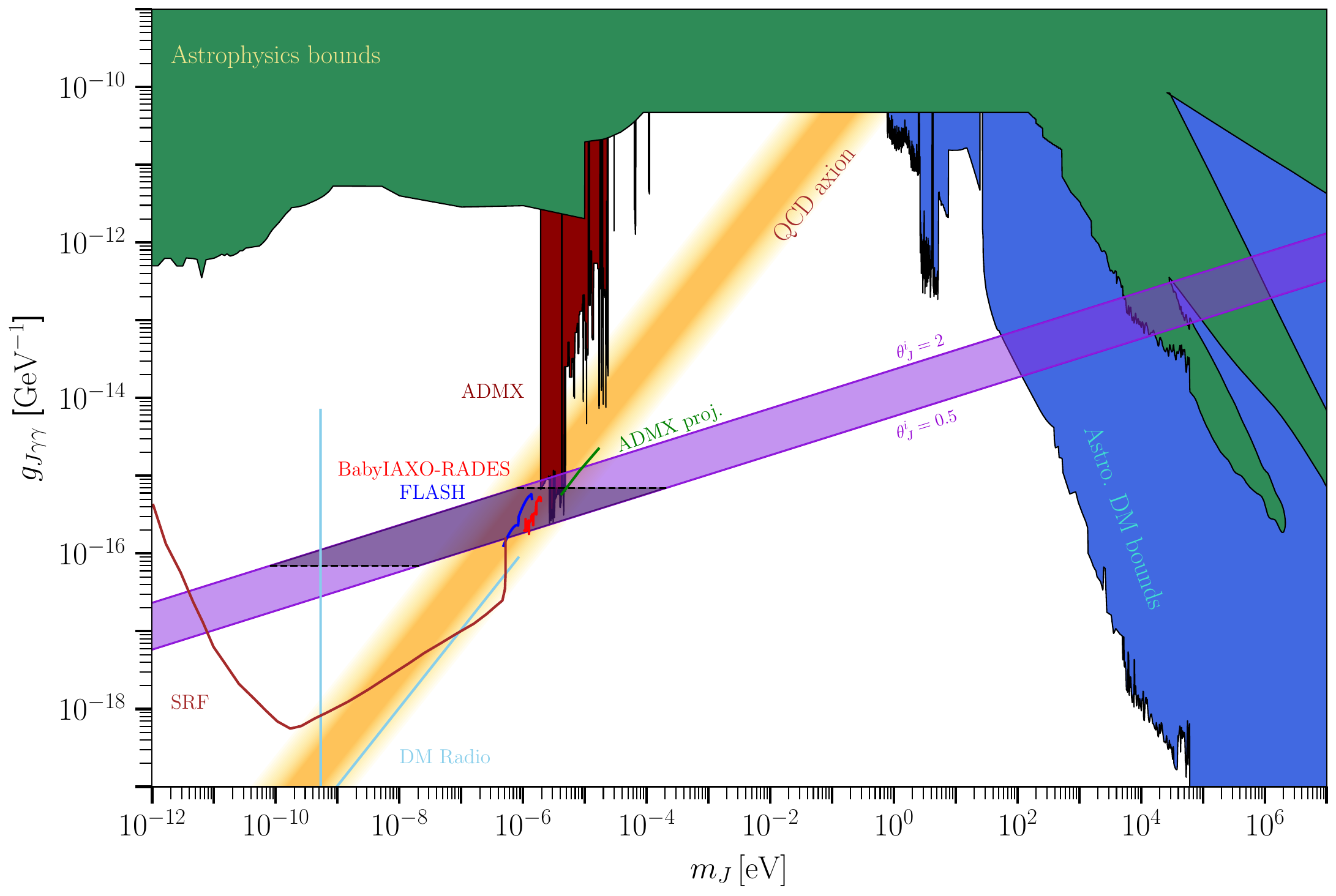}
    \caption{In purple we see the prediction from misalignment for our majoron. The preferred (black) region is determined by varying the initial misalignment angle $\theta_J^i\in\left[0.5,\, 2\right]$, and of $F_J\in \left[10^{13},\,10^{14}\right]\, \textrm{GeV}$.}
    \label{fig:MajoronDarkMatter}
\end{figure*}
\noindent where $\theta_J^i\sim \mathcal{O}(1)$ is the initial angle at the start of the oscillations \footnote {We assume that the lepton-number breaking occurs during or before the inflation. The scale of inflation $H_{inf}$ needs to be sufficiently small to avoid large isocurvature fluctuations \cite{Bardeen:1983qw,Seckel:1985tj}.}. 
This sets the scale-mass relation of the majoron DM. Considering the natural scale for the seesaw and leptogenesis $F_J\sim 10^{13}-10^{14}\,$GeV \cite{Buchmuller:2005eh}, the majoron mass are natural light. Hence, traditional methods to detect majoron would fail. However, the model in Eq.~\eqref{eq:Lagrangian1} presents an anomalous coupling to QED, whose coupling constant can be related to the majoron mass through,
\begin{eqnarray}
    m_J \simeq \left(\frac{\pi}{3\alpha_{\text{em}} \theta^{i }_J   } \frac{g_{J\gamma\gamma}}{ 1.9\times 10^{-13} \text{GeV}}\right)^{4}\mu \textrm{eV} \ ,
\end{eqnarray}
which is optimal for detection as we will discuss now.

In Fig.~\ref{fig:MajoronDarkMatter} we plot the current constraints on the anomalous coupling, including both astrophysical, cosmological, and experimental bounds \cite{AxionLimits}. We further plot the the prediction of $g_{J\gamma\gamma}$ in our model with mass $m_J\in\left[10^{-12},\, 10^{6}\right]$ eV in purple, with a variation of initial angle $\theta_J^i\in\left[0.5,\, 2\right]$. The favoured region by seesaw mechanism and leptogenesis are coloured in black. Surprisingly, axion experiments meet the majoron band! In particular, ADMX experiment \cite{ADMX:2018gho,ADMX:2019uok,ADMX:2021nhd} has already reached the preferred band for this model. 

In the future, there exists several proposals for axion experiments that will explore further the preferred region for this majoron. In Fig.~\ref{fig:MajoronDarkMatter} we see several projections that can test both particles, the QCD axion and the majoron. Firstly, we have the future searches of  ADMX \footnote{See, talk at \href{https://indico.cern.ch/event/1199289/contributions/5449605/attachments/2705162/4697491/TAUP2023_Oblath_ADMX_20231830.pdf}{TAUP23}.}, sweeping the region of $\mathcal{O}(1-10)\mu$eV. Then, for lower masses, the proposals of implementing a haloscope inside BabyIAXO's magnet, RADES \cite{Ahyoune:2023gfw}, and the proposal of using FINUDA's magnet, FLASH \cite{Alesini:2023qed}, could investigate the majoron region of $\mathcal{O}(0.1)\mu$eV. Below these masses, the method of testing the $J F\tilde F$-term using resonant cavities becomes unfeasible, as the length of the cavity needs to be comparable to the Compton wavelength of the majoron. However, one can still test this low regions using Superconducting Radio Frequency (SRF) cavities \cite{Berlin:2020vrk}, or the DM-radio experiment \cite{DMRadio:2022pkf} which can detect the magnetic field generated by the majoron. All in all we see, this particle is testable for the preferred region of the model with the current proposals of axion detection.

\section{Conclusions and discussion}
\label{sec:conclusions}
The majoron is a well-motivated light boson which is generated by spontaneous breaking of a global lepton-number $U(1)_L$ symmetry. The symmetry breaking induces large Majorana masses for the right-handed neutrinos, and at the same time a massless Nambu-Goldstone boson which we call majoron $J$. By introducing a second Higgs field in the model, we can assign different lepton charges to obtain an anomalous coupling $g_{J\gamma\gamma}$ under QED. 

On the other hand, the majoron can also serve as dark matter. The soft-breaking mass $m_J$ is introduced so that the majoron coherent oscillation explains the observed DM density which depends on the lepton-number breaking scale $v_\phi $. Combining our prediction of the $g_{J\gamma\gamma}$, we have plotted the prediction of the majoron DM hypothesis in Fig.~\ref{fig:MajoronDarkMatter}. Surprisingly, the region of the majoron DM overlaps largely the prediction of the axion DM, seeing how a large set of experimental proposals searching for the QCD axion will also meet the majoron in the future!

It might be interesting to assume the global symmetry is exact beside gauge anomaly terms. In fact, our $U(1)_L$ global current has $SU(2)\times SU(2)$ anomaly. Then, the majoron mass can be induced by the electroweak instantons. However, we immediately realize the obtained mass is too small to be the dark matter due to the small gauge coupling constant at small scales. Nonetheless, there is an intriguing possibility to identify the majoron with the quintessence axion \cite{Nomura:2000yk,Choi:2021aze} or the electroweak axion \cite{Lin:2022niw} responsible for the cosmic birefrigence \cite{Minami:2020odp,Diego-Palazuelos:2022dsq}.
 
We would, finally, remark that the two heavy Majorana right-handed neutrinos are sufficient \cite{Frampton:2002qc} for generating the baryon-number asymmetry in the universe. If it is the case, the lightest neutrino is exactly massless in our framework. The presence of the two right-handed neutrinos is consistent with our global lepton-number $U(1)_L$ symmetry.

\section*{Ackowledgments}

This work is supported by 
JSPS Grant-in-Aid for Scientific Research
Grants No.\,24H02244, the National Natural Science
Foundation of China (12175134)
and World Premier International Research Center
Initiative (WPI Initiative), MEXT, Japan. This project has received funding from the European Union’s
Horizon 2020 research and innovation programme under the Marie Skłodowska-Curie
grant agreement N. 860881-HIDDeN. XPD thanks Kavli IPMU for its hospitality during the development of this work. We thank Ciaran O'Hare for useful comments on experimental projections, and Luca Di Luzio for useful comments on the draft.

\appendix
\section{Details}
\label{sec:appendix}

There are two possible lepton number charge assignments, given by the exchange of the two Higgs doublets in the lepton sector. The first one is anomalous under QED
\begin{align}
    \label{eq:LN1lag}
    \mathcal{L}&= \bar{q}_L  Y_u u_R \tilde{H}_1 + \bar{q}_L  Y_d d_R H_1 +\bar{\ell}_L Y_e e_R H_2\nonumber \\ & +\bar{\ell}_L Y_e N_R \tilde{H}_1 + \bar{N}_R^c N_R \phi^* \nonumber +H_2^\dagger H_1 \phi \\ & +V(H_1,\,H_2,\, \phi)+\textrm{h.c.}\, ,
\end{align}
the second is not anomalous under QED,
\begin{align}
    \mathcal{L} &= \bar{q}_L  Y_u u_R \tilde{H}_1 + \bar{q}_L  Y_d d_R H_1 +\bar{\ell}_L Y_e e_R H_1 \nonumber\\ &+\bar{\ell}_L Y_e N_R \tilde{H}_2 + \bar{N}_R^c N_R \phi^* +H_2^\dagger H_1 \phi \nonumber \\ & +V(H_1,\,H_2,\, \phi)+\textrm{h.c.}\, . 
\end{align}
Here $H_1$, $H_2$ are $SU(2)_L$ Higgs doublets and $\phi$ is the complex scalar that breaks spontaneously the LN symmetry. There is a global symmetry defined by the interactions of each Lagrangian, with the charges $\X$ given in Tab.~\ref{tab:particle_content}. 

The majoron can be defined in this case by the current associated, following Goldstone's theorem $\bra{0} J_\mu \ket{J(k)}=-i F_J k_\mu $,
\begin{equation}
    J = \frac{1}{F_J} \sum_i \X_i v_i a_i\, ,    \textrm{ with } F_J=\sum_i \X_i^2 v_i^2 \,,
\end{equation}
where $a_i$ are the phases of the scalars $H_i\supset \exp(i a_i/v_i)$, $\phi\supset \exp(i a_\phi/v_\phi)$, and $v_i$ the different vevs. In this \textit{letter}, we take $v_\phi\gg v_i$, so that the majoron can be, in practice, associated to $J\simeq a_\phi$ with $F_J\simeq v_\phi$.
\begin{table*}[tb]
\centering
\begin{tabular}{|c|c|c|c|c|c|}
\hline
Fields & $SU(3)_c$ & $SU(2)_L$ & $U(1)_Y$ & LN 1 & LN 2 \\ \hline
$q_L$ & 3 & 2 & 1/6 & $\X_q$ & $\X_q$\\ 
$u_R$ & 3 & 1 & 2/3 & $\X_q+\X_1$ & $\X_q+\X_1$ \\ 
$d_R$ & 3 & 1 & -1/3 & $\X_q-\X_1$ & $\X_q-\X_1$\\ 
$\ell_L$ & 1 & 2 & -1/2 & $1/2-\X_1$ & $-3/2-\X_1$\\ 
$e_R$ & 1 & 1 & -1 & $-1/2-2\X_1$ & $-3/2-2\X_1$\\ 
$N_R$ & 1 & 1 & 0 & 1/2  & 1/2\\ \hline 
$H_1$ & 1 & 2 & 1/2 &   $\X_{1}$& $\X_{1}$\\
$H_2$ & 1 & 2 & 1/2 & $1+\X_1$  &  $1+\X_1$\\
$\phi$ & 1 & 1 & 0 & 1 &1\\ \hline
\end{tabular}
\caption{Particle content and charges of the particles with respect to the gauge group of the SM and Lepton Number of the two models, LN1 is the model presented in this work, while LN2 is a possible different non-anomalous set of chages.}
\label{tab:particle_content}
\end{table*}

From the table we can compute the anomalous coefficient to these charge assignments. The anomalous terms are defined by the Lagrangian
\begin{equation}
    \mathcal{L}_{\textrm{Anom.}} = \frac{\alpha_{\textrm{em}}}{4 \pi} E \frac{J}{F_J} F_{\mu\nu}\widetilde{F}^{\mu\nu}+\frac{\alpha_{s}}{4 \pi} N \frac{J}{F_J}G^a_{\mu\nu}\widetilde{G}^{a\,\mu\nu}\, ,
\end{equation}
where for the first model they can be computed as
\begin{align}   
    N &= \frac{n_f}{2} \left(2\X_q-\X_u-\X_d\right) =\frac{n_f}{2} \left(\X_1-\X_1\right) =0 \, ,\nonumber \\
    E &= n_f \sum_i \X_i Q^2_i = n_f\left(\frac{4}{3}\X_q+\frac{1}{3}\X_q-\frac{4}{3}\X_u \right.\nonumber \\ &\left. -\frac{1}{3}\X_d+(\X_\ell-\X_{e})\right)= n_f\left(-\X_1+\X_2 \right)\nonumber \\
    &= n_f\times \X_\phi  =3  \, ,
\end{align}
where $Q_i$ are the QED charges of the SM fermions and $\X_\phi=1$.

Noting that there is no QCD anomaly and that the electromagnetic anomaly is independent of the choice of $\X_{1,2}$, in the main text we have taken $\X_1 =0$ and $\X_q=0$, for simplicity.
However, it is important to note that this choice generates a mixing with the $Z$-boson coming from the kinetic term
\begin{align}
    \mathcal{L}_{kin.}&\supset (D_\mu H_1)^\dagger D_\mu H_1+(D_\mu H_2)^\dagger D_\mu H_2 \nonumber \\&\supset \frac{\partial_\mu J}{F_J} Z^\mu \frac{g'}{2}\left(\X_1 v_1^2 +\X_2 v_2^2\right)\, .
\end{align}
It is convenient to avoid such a mixing, defining the charges as $\X_1=-\sin^2\beta$ and $\X_2=\cos^2\beta$, with $\tan\beta=v_2/v_1$. This choice, inevitably leads to charging the quark sector, having the interactions
\begin{align}
    \mathcal{L}_{J}^{\textrm{SM}}=&\frac{\partial^\mu J}{2F_J} \left(-\sin^2\beta\bar u \gamma_\mu\gamma_5 u+\sin^2\beta\bar d \gamma_\mu\gamma_5 d \right.\nonumber\\ &\left.-\cos^2\beta\bar e \gamma_\mu\gamma_5 e \right)+3\frac{\alpha_{em}}{4\pi} \frac{J}{F_J}F\tilde F \, .
\end{align}
The reason why no emphasis has been given in the main text to this couplings is that for such small masses of the majoron and large scale, these couplings are rather difficult to test and no bounds are applied.

The neutrino sector is then given by
\begin{equation}
    \mathcal{L}_{J}^\nu \simeq \frac{\partial^\mu J}{2F_J}\left(-\frac{1}{2}\bar{\nu} \gamma_\mu \gamma_5 \nu + \frac{1}{2}\bar N \gamma_\mu \gamma_5 N\right)
\end{equation}
with $\nu=\nu_L+\nu_L^c$ and similar for $N$. This formula is approximate, as the diagonalization of the neutrino mass matrix induces $\sim \mathcal{O}(v/v_\phi)$ corrections, that we can safely neglect.

In Eq.~\eqref{eq:LN1lag}, we have only mentioned the coupling $H_2^\dagger H_1\phi$ of the potential, which is enough for setting the charges. Other options exist, such as $H_2^\dagger H_1\phi^2$, which would give an equally valid model with $E=6$. Finally, it is worth noting that the extra scalar degrees of freedom can be decoupled from the theory. This discussion is analogous to axion models, see for instance Appendix A of Ref.~\cite{DiLuzio:2023ndz} for details.

\bibliography{bibliography}

\begin{thebibliography}{51}%
\makeatletter
\providecommand \@ifxundefined [1]{%
 \@ifx{#1\undefined}
}%
\providecommand \@ifnum [1]{%
 \ifnum #1\expandafter \@firstoftwo
 \else \expandafter \@secondoftwo
 \fi
}%
\providecommand \@ifx [1]{%
 \ifx #1\expandafter \@firstoftwo
 \else \expandafter \@secondoftwo
 \fi
}%
\providecommand \natexlab [1]{#1}%
\providecommand \enquote  [1]{``#1''}%
\providecommand \bibnamefont  [1]{#1}%
\providecommand \bibfnamefont [1]{#1}%
\providecommand \citenamefont [1]{#1}%
\providecommand \href@noop [0]{\@secondoftwo}%
\providecommand \href [0]{\begingroup \@sanitize@url \@href}%
\providecommand \@href[1]{\@@startlink{#1}\@@href}%
\providecommand \@@href[1]{\endgroup#1\@@endlink}%
\providecommand \@sanitize@url [0]{\catcode `\\12\catcode `\$12\catcode
  `\&12\catcode `\#12\catcode `\^12\catcode `\_12\catcode `\%12\relax}%
\providecommand \@@startlink[1]{}%
\providecommand \@@endlink[0]{}%
\providecommand \url  [0]{\begingroup\@sanitize@url \@url }%
\providecommand \@url [1]{\endgroup\@href {#1}{\urlprefix }}%
\providecommand \urlprefix  [0]{URL }%
\providecommand \Eprint [0]{\href }%
\providecommand \doibase [0]{https://doi.org/}%
\providecommand \selectlanguage [0]{\@gobble}%
\providecommand \bibinfo  [0]{\@secondoftwo}%
\providecommand \bibfield  [0]{\@secondoftwo}%
\providecommand \translation [1]{[#1]}%
\providecommand \BibitemOpen [0]{}%
\providecommand \bibitemStop [0]{}%
\providecommand \bibitemNoStop [0]{.\EOS\space}%
\providecommand \EOS [0]{\spacefactor3000\relax}%
\providecommand \BibitemShut  [1]{\csname bibitem#1\endcsname}%
\let\auto@bib@innerbib\@empty
%</preamble>
\bibitem [{\citenamefont {Minkowski}(1977)}]{Minkowski:1977sc}%
  \BibitemOpen
  \bibfield  {author} {\bibinfo {author} {\bibfnamefont {P.}~\bibnamefont
  {Minkowski}},\ }\bibfield  {title} {\bibinfo {title} {{$\mu \to e\gamma$ at a
  Rate of One Out of $10^{9}$ Muon Decays?}},\ }\href
  {https://doi.org/10.1016/0370-2693(77)90435-X} {\bibfield  {journal}
  {\bibinfo  {journal} {Phys. Lett. B}\ }\textbf {\bibinfo {volume} {67}},\
  \bibinfo {pages} {421} (\bibinfo {year} {1977})}\BibitemShut {NoStop}%
\bibitem [{\citenamefont {Yanagida}(1979{\natexlab{a}})}]{Yanagida:1979as}%
  \BibitemOpen
  \bibfield  {author} {\bibinfo {author} {\bibfnamefont {T.}~\bibnamefont
  {Yanagida}},\ }\bibfield  {title} {\bibinfo {title} {{Horizontal gauge
  symmetry and masses of neutrinos}},\ }\href@noop {} {\bibfield  {journal}
  {\bibinfo  {journal} {Conf. Proc. C}\ }\textbf {\bibinfo {volume}
  {7902131}},\ \bibinfo {pages} {95} (\bibinfo {year}
  {1979}{\natexlab{a}})}\BibitemShut {NoStop}%
\bibitem [{\citenamefont {Yanagida}(1979{\natexlab{b}})}]{Yanagida:1979gs}%
  \BibitemOpen
  \bibfield  {author} {\bibinfo {author} {\bibfnamefont {T.}~\bibnamefont
  {Yanagida}},\ }\bibfield  {title} {\bibinfo {title} {{Horizontal Symmetry and
  Mass of the Top Quark}},\ }\href {https://doi.org/10.1103/PhysRevD.20.2986}
  {\bibfield  {journal} {\bibinfo  {journal} {Phys. Rev. D}\ }\textbf {\bibinfo
  {volume} {20}},\ \bibinfo {pages} {2986} (\bibinfo {year}
  {1979}{\natexlab{b}})}\BibitemShut {NoStop}%
\bibitem [{\citenamefont {Glashow}(1980)}]{Glashow:1979nm}%
  \BibitemOpen
  \bibfield  {author} {\bibinfo {author} {\bibfnamefont {S.~L.}\ \bibnamefont
  {Glashow}},\ }\bibfield  {title} {\bibinfo {title} {{The Future of Elementary
  Particle Physics}},\ }\href {https://doi.org/10.1007/978-1-4684-7197-7_15}
  {\bibfield  {journal} {\bibinfo  {journal} {NATO Sci. Ser. B}\ }\textbf
  {\bibinfo {volume} {61}},\ \bibinfo {pages} {687} (\bibinfo {year}
  {1980})}\BibitemShut {NoStop}%
\bibitem [{\citenamefont {Gell-Mann}\ \emph {et~al.}(1979)\citenamefont
  {Gell-Mann}, \citenamefont {Ramond},\ and\ \citenamefont
  {Slansky}}]{Gell-Mann:1979vob}%
  \BibitemOpen
  \bibfield  {author} {\bibinfo {author} {\bibfnamefont {M.}~\bibnamefont
  {Gell-Mann}}, \bibinfo {author} {\bibfnamefont {P.}~\bibnamefont {Ramond}},\
  and\ \bibinfo {author} {\bibfnamefont {R.}~\bibnamefont {Slansky}},\
  }\bibfield  {title} {\bibinfo {title} {{Complex Spinors and Unified
  Theories}},\ }\href@noop {} {\bibfield  {journal} {\bibinfo  {journal} {Conf.
  Proc. C}\ }\textbf {\bibinfo {volume} {790927}},\ \bibinfo {pages} {315}
  (\bibinfo {year} {1979})},\ \Eprint {https://arxiv.org/abs/1306.4669}
  {arXiv:1306.4669 [hep-th]} \BibitemShut {NoStop}%
\bibitem [{\citenamefont {Weinberg}(1979)}]{Weinberg:1979sa}%
  \BibitemOpen
  \bibfield  {author} {\bibinfo {author} {\bibfnamefont {S.}~\bibnamefont
  {Weinberg}},\ }\bibfield  {title} {\bibinfo {title} {{Baryon and Lepton
  Nonconserving Processes}},\ }\href
  {https://doi.org/10.1103/PhysRevLett.43.1566} {\bibfield  {journal} {\bibinfo
   {journal} {Phys. Rev. Lett.}\ }\textbf {\bibinfo {volume} {43}},\ \bibinfo
  {pages} {1566} (\bibinfo {year} {1979})}\BibitemShut {NoStop}%
\bibitem [{\citenamefont {Wilczeck}(1979)}]{Wilczeck:1979CP}%
  \BibitemOpen
  \bibfield  {author} {\bibinfo {author} {\bibfnamefont {F.}~\bibnamefont
  {Wilczeck}},\ }\href@noop {} {\bibfield  {journal} {\bibinfo  {journal}
  {{\textit{Proceedings: Lepton-Photon Conference (Fermilab, Aug 1979)}} Conf.
  Proc.}\ }\textbf {\bibinfo {volume} {C790885}} (\bibinfo {year}
  {1979})}\BibitemShut {NoStop}%
\bibitem [{\citenamefont {Barbieri}\ \emph {et~al.}(1980)\citenamefont
  {Barbieri}, \citenamefont {Nanopoulos}, \citenamefont {Morchio},\ and\
  \citenamefont {Strocchi}}]{Barbieri:1979ag}%
  \BibitemOpen
  \bibfield  {author} {\bibinfo {author} {\bibfnamefont {R.}~\bibnamefont
  {Barbieri}}, \bibinfo {author} {\bibfnamefont {D.~V.}\ \bibnamefont
  {Nanopoulos}}, \bibinfo {author} {\bibfnamefont {G.}~\bibnamefont
  {Morchio}},\ and\ \bibinfo {author} {\bibfnamefont {F.}~\bibnamefont
  {Strocchi}},\ }\bibfield  {title} {\bibinfo {title} {{Neutrino Masses in
  Grand Unified Theories}},\ }\href
  {https://doi.org/10.1016/0370-2693(80)90058-1} {\bibfield  {journal}
  {\bibinfo  {journal} {Phys. Lett. B}\ }\textbf {\bibinfo {volume} {90}},\
  \bibinfo {pages} {91} (\bibinfo {year} {1980})}\BibitemShut {NoStop}%
\bibitem [{\citenamefont {Witten}(1980)}]{Witten:1979nr}%
  \BibitemOpen
  \bibfield  {author} {\bibinfo {author} {\bibfnamefont {E.}~\bibnamefont
  {Witten}},\ }\bibfield  {title} {\bibinfo {title} {{Neutrino Masses in the
  Minimal O(10) Theory}},\ }\href
  {https://doi.org/10.1016/0370-2693(80)90666-8} {\bibfield  {journal}
  {\bibinfo  {journal} {Phys. Lett. B}\ }\textbf {\bibinfo {volume} {91}},\
  \bibinfo {pages} {81} (\bibinfo {year} {1980})}\BibitemShut {NoStop}%
\bibitem [{\citenamefont {Mohapatra}\ and\ \citenamefont
  {Senjanovic}(1980)}]{Mohapatra:1979ia}%
  \BibitemOpen
  \bibfield  {author} {\bibinfo {author} {\bibfnamefont {R.~N.}\ \bibnamefont
  {Mohapatra}}\ and\ \bibinfo {author} {\bibfnamefont {G.}~\bibnamefont
  {Senjanovic}},\ }\bibfield  {title} {\bibinfo {title} {{Neutrino Mass and
  Spontaneous Parity Nonconservation}},\ }\href
  {https://doi.org/10.1103/PhysRevLett.44.912} {\bibfield  {journal} {\bibinfo
  {journal} {Phys. Rev. Lett.}\ }\textbf {\bibinfo {volume} {44}},\ \bibinfo
  {pages} {912} (\bibinfo {year} {1980})}\BibitemShut {NoStop}%
\bibitem [{\citenamefont {Fukugita}\ and\ \citenamefont
  {Yanagida}(1986)}]{Fukugita:1986hr}%
  \BibitemOpen
  \bibfield  {author} {\bibinfo {author} {\bibfnamefont {M.}~\bibnamefont
  {Fukugita}}\ and\ \bibinfo {author} {\bibfnamefont {T.}~\bibnamefont
  {Yanagida}},\ }\bibfield  {title} {\bibinfo {title} {{Baryogenesis Without
  Grand Unification}},\ }\href {https://doi.org/10.1016/0370-2693(86)91126-3}
  {\bibfield  {journal} {\bibinfo  {journal} {Phys. Lett. B}\ }\textbf
  {\bibinfo {volume} {174}},\ \bibinfo {pages} {45} (\bibinfo {year}
  {1986})}\BibitemShut {NoStop}%
\bibitem [{\citenamefont {Buchmuller}\ \emph {et~al.}(2005)\citenamefont
  {Buchmuller}, \citenamefont {Peccei},\ and\ \citenamefont
  {Yanagida}}]{Buchmuller:2005eh}%
  \BibitemOpen
  \bibfield  {author} {\bibinfo {author} {\bibfnamefont {W.}~\bibnamefont
  {Buchmuller}}, \bibinfo {author} {\bibfnamefont {R.~D.}\ \bibnamefont
  {Peccei}},\ and\ \bibinfo {author} {\bibfnamefont {T.}~\bibnamefont
  {Yanagida}},\ }\bibfield  {title} {\bibinfo {title} {{Leptogenesis as the
  origin of matter}},\ }\href
  {https://doi.org/10.1146/annurev.nucl.55.090704.151558} {\bibfield  {journal}
  {\bibinfo  {journal} {Ann. Rev. Nucl. Part. Sci.}\ }\textbf {\bibinfo
  {volume} {55}},\ \bibinfo {pages} {311} (\bibinfo {year} {2005})},\ \Eprint
  {https://arxiv.org/abs/hep-ph/0502169} {arXiv:hep-ph/0502169} \BibitemShut
  {NoStop}%
\bibitem [{\citenamefont {Chikashige}\ \emph {et~al.}(1980)\citenamefont
  {Chikashige}, \citenamefont {Mohapatra},\ and\ \citenamefont
  {Peccei}}]{Chikashige:1980qk}%
  \BibitemOpen
  \bibfield  {author} {\bibinfo {author} {\bibfnamefont {Y.}~\bibnamefont
  {Chikashige}}, \bibinfo {author} {\bibfnamefont {R.~N.}\ \bibnamefont
  {Mohapatra}},\ and\ \bibinfo {author} {\bibfnamefont {R.~D.}\ \bibnamefont
  {Peccei}},\ }\bibfield  {title} {\bibinfo {title} {{Spontaneously Broken
  Lepton Number and Cosmological Constraints on the Neutrino Mass Spectrum}},\
  }\href {https://doi.org/10.1103/PhysRevLett.45.1926} {\bibfield  {journal}
  {\bibinfo  {journal} {Phys. Rev. Lett.}\ }\textbf {\bibinfo {volume} {45}},\
  \bibinfo {pages} {1926} (\bibinfo {year} {1980})}\BibitemShut {NoStop}%
\bibitem [{\citenamefont {Chikashige}\ \emph {et~al.}(1981)\citenamefont
  {Chikashige}, \citenamefont {Mohapatra},\ and\ \citenamefont
  {Peccei}}]{Chikashige:1980ui}%
  \BibitemOpen
  \bibfield  {author} {\bibinfo {author} {\bibfnamefont {Y.}~\bibnamefont
  {Chikashige}}, \bibinfo {author} {\bibfnamefont {R.~N.}\ \bibnamefont
  {Mohapatra}},\ and\ \bibinfo {author} {\bibfnamefont {R.~D.}\ \bibnamefont
  {Peccei}},\ }\bibfield  {title} {\bibinfo {title} {{Are There Real Goldstone
  Bosons Associated with Broken Lepton Number?}},\ }\href
  {https://doi.org/10.1016/0370-2693(81)90011-3} {\bibfield  {journal}
  {\bibinfo  {journal} {Phys. Lett. B}\ }\textbf {\bibinfo {volume} {98}},\
  \bibinfo {pages} {265} (\bibinfo {year} {1981})}\BibitemShut {NoStop}%
\bibitem [{\citenamefont {Gelmini}\ and\ \citenamefont
  {Roncadelli}(1981)}]{Gelmini:1980re}%
  \BibitemOpen
  \bibfield  {author} {\bibinfo {author} {\bibfnamefont {G.~B.}\ \bibnamefont
  {Gelmini}}\ and\ \bibinfo {author} {\bibfnamefont {M.}~\bibnamefont
  {Roncadelli}},\ }\bibfield  {title} {\bibinfo {title} {{Left-Handed Neutrino
  Mass Scale and Spontaneously Broken Lepton Number}},\ }\href
  {https://doi.org/10.1016/0370-2693(81)90559-1} {\bibfield  {journal}
  {\bibinfo  {journal} {Phys. Lett. B}\ }\textbf {\bibinfo {volume} {99}},\
  \bibinfo {pages} {411} (\bibinfo {year} {1981})}\BibitemShut {NoStop}%
\bibitem [{\citenamefont {Gu}\ \emph {et~al.}(2010)\citenamefont {Gu},
  \citenamefont {Ma},\ and\ \citenamefont {Sarkar}}]{Gu:2010ys}%
  \BibitemOpen
  \bibfield  {author} {\bibinfo {author} {\bibfnamefont {P.-H.}\ \bibnamefont
  {Gu}}, \bibinfo {author} {\bibfnamefont {E.}~\bibnamefont {Ma}},\ and\
  \bibinfo {author} {\bibfnamefont {U.}~\bibnamefont {Sarkar}},\ }\bibfield
  {title} {\bibinfo {title} {{Pseudo-Majoron as Dark Matter}},\ }\href
  {https://doi.org/10.1016/j.physletb.2010.05.012} {\bibfield  {journal}
  {\bibinfo  {journal} {Phys. Lett. B}\ }\textbf {\bibinfo {volume} {690}},\
  \bibinfo {pages} {145} (\bibinfo {year} {2010})},\ \Eprint
  {https://arxiv.org/abs/1004.1919} {arXiv:1004.1919 [hep-ph]} \BibitemShut
  {NoStop}%
\bibitem [{\citenamefont {Frigerio}\ \emph {et~al.}(2011)\citenamefont
  {Frigerio}, \citenamefont {Hambye},\ and\ \citenamefont
  {Masso}}]{Frigerio:2011in}%
  \BibitemOpen
  \bibfield  {author} {\bibinfo {author} {\bibfnamefont {M.}~\bibnamefont
  {Frigerio}}, \bibinfo {author} {\bibfnamefont {T.}~\bibnamefont {Hambye}},\
  and\ \bibinfo {author} {\bibfnamefont {E.}~\bibnamefont {Masso}},\ }\bibfield
   {title} {\bibinfo {title} {{Sub-GeV dark matter as pseudo-Goldstone from the
  seesaw scale}},\ }\href {https://doi.org/10.1103/PhysRevX.1.021026}
  {\bibfield  {journal} {\bibinfo  {journal} {Phys. Rev. X}\ }\textbf {\bibinfo
  {volume} {1}},\ \bibinfo {pages} {021026} (\bibinfo {year} {2011})},\ \Eprint
  {https://arxiv.org/abs/1107.4564} {arXiv:1107.4564 [hep-ph]} \BibitemShut
  {NoStop}%
\bibitem [{\citenamefont {Giddings}\ and\ \citenamefont
  {Strominger}(1988)}]{Giddings:1987cg}%
  \BibitemOpen
  \bibfield  {author} {\bibinfo {author} {\bibfnamefont {S.~B.}\ \bibnamefont
  {Giddings}}\ and\ \bibinfo {author} {\bibfnamefont {A.}~\bibnamefont
  {Strominger}},\ }\bibfield  {title} {\bibinfo {title} {{Axion Induced
  Topology Change in Quantum Gravity and String Theory}},\ }\href
  {https://doi.org/10.1016/0550-3213(88)90446-4} {\bibfield  {journal}
  {\bibinfo  {journal} {Nucl. Phys. B}\ }\textbf {\bibinfo {volume} {306}},\
  \bibinfo {pages} {890} (\bibinfo {year} {1988})}\BibitemShut {NoStop}%
\bibitem [{\citenamefont {Alonso}\ and\ \citenamefont
  {Urbano}(2019)}]{Alonso:2017avz}%
  \BibitemOpen
  \bibfield  {author} {\bibinfo {author} {\bibfnamefont {R.}~\bibnamefont
  {Alonso}}\ and\ \bibinfo {author} {\bibfnamefont {A.}~\bibnamefont
  {Urbano}},\ }\bibfield  {title} {\bibinfo {title} {{Wormholes and masses for
  Goldstone bosons}},\ }\href {https://doi.org/10.1007/JHEP02(2019)136}
  {\bibfield  {journal} {\bibinfo  {journal} {JHEP}\ }\textbf {\bibinfo
  {volume} {02}},\ \bibinfo {pages} {136}},\ \Eprint
  {https://arxiv.org/abs/1706.07415} {arXiv:1706.07415 [hep-ph]} \BibitemShut
  {NoStop}%
\bibitem [{\citenamefont {Coleman}(1985)}]{Coleman:1985rnk}%
  \BibitemOpen
  \bibfield  {author} {\bibinfo {author} {\bibfnamefont {S.}~\bibnamefont
  {Coleman}},\ }\href {https://doi.org/10.1017/CBO9780511565045} {\emph
  {\bibinfo {title} {{Aspects of Symmetry}: {Selected Erice Lectures}}}}\
  (\bibinfo  {publisher} {Cambridge University Press},\ \bibinfo {address}
  {Cambridge, U.K.},\ \bibinfo {year} {1985})\BibitemShut {NoStop}%
\bibitem [{\citenamefont {Abbott}\ and\ \citenamefont
  {Sikivie}(1983)}]{Abbott:1982af}%
  \BibitemOpen
  \bibfield  {author} {\bibinfo {author} {\bibfnamefont {L.~F.}\ \bibnamefont
  {Abbott}}\ and\ \bibinfo {author} {\bibfnamefont {P.}~\bibnamefont
  {Sikivie}},\ }\bibfield  {title} {\bibinfo {title} {{A Cosmological Bound on
  the Invisible Axion}},\ }\href {https://doi.org/10.1016/0370-2693(83)90638-X}
  {\bibfield  {journal} {\bibinfo  {journal} {Phys. Lett. B}\ }\textbf
  {\bibinfo {volume} {120}},\ \bibinfo {pages} {133} (\bibinfo {year}
  {1983})}\BibitemShut {NoStop}%
\bibitem [{\citenamefont {Dine}\ \emph {et~al.}(1981)\citenamefont {Dine},
  \citenamefont {Fischler},\ and\ \citenamefont {Srednicki}}]{Dine:1981rt}%
  \BibitemOpen
  \bibfield  {author} {\bibinfo {author} {\bibfnamefont {M.}~\bibnamefont
  {Dine}}, \bibinfo {author} {\bibfnamefont {W.}~\bibnamefont {Fischler}},\
  and\ \bibinfo {author} {\bibfnamefont {M.}~\bibnamefont {Srednicki}},\
  }\bibfield  {title} {\bibinfo {title} {{A Simple Solution to the Strong CP
  Problem with a Harmless Axion}},\ }\href
  {https://doi.org/10.1016/0370-2693(81)90590-6} {\bibfield  {journal}
  {\bibinfo  {journal} {Phys. Lett. B}\ }\textbf {\bibinfo {volume} {104}},\
  \bibinfo {pages} {199} (\bibinfo {year} {1981})}\BibitemShut {NoStop}%
\bibitem [{\citenamefont {Preskill}\ \emph {et~al.}(1983)\citenamefont
  {Preskill}, \citenamefont {Wise},\ and\ \citenamefont
  {Wilczek}}]{Preskill:1982cy}%
  \BibitemOpen
  \bibfield  {author} {\bibinfo {author} {\bibfnamefont {J.}~\bibnamefont
  {Preskill}}, \bibinfo {author} {\bibfnamefont {M.~B.}\ \bibnamefont {Wise}},\
  and\ \bibinfo {author} {\bibfnamefont {F.}~\bibnamefont {Wilczek}},\
  }\bibfield  {title} {\bibinfo {title} {{Cosmology of the Invisible Axion}},\
  }\href {https://doi.org/10.1016/0370-2693(83)90637-8} {\bibfield  {journal}
  {\bibinfo  {journal} {Phys. Lett. B}\ }\textbf {\bibinfo {volume} {120}},\
  \bibinfo {pages} {127} (\bibinfo {year} {1983})}\BibitemShut {NoStop}%
\bibitem [{\citenamefont {Garcia-Cely}\ and\ \citenamefont
  {Heeck}(2017)}]{Garcia-Cely:2017oco}%
  \BibitemOpen
  \bibfield  {author} {\bibinfo {author} {\bibfnamefont {C.}~\bibnamefont
  {Garcia-Cely}}\ and\ \bibinfo {author} {\bibfnamefont {J.}~\bibnamefont
  {Heeck}},\ }\bibfield  {title} {\bibinfo {title} {{Neutrino Lines from
  Majoron Dark Matter}},\ }\href {https://doi.org/10.1007/JHEP05(2017)102}
  {\bibfield  {journal} {\bibinfo  {journal} {JHEP}\ }\textbf {\bibinfo
  {volume} {05}},\ \bibinfo {pages} {102}},\ \Eprint
  {https://arxiv.org/abs/1701.07209} {arXiv:1701.07209 [hep-ph]} \BibitemShut
  {NoStop}%
\bibitem [{\citenamefont {Akita}\ and\ \citenamefont
  {Niibo}(2023)}]{Akita:2023qiz}%
  \BibitemOpen
  \bibfield  {author} {\bibinfo {author} {\bibfnamefont {K.}~\bibnamefont
  {Akita}}\ and\ \bibinfo {author} {\bibfnamefont {M.}~\bibnamefont {Niibo}},\
  }\bibfield  {title} {\bibinfo {title} {{Updated constraints and future
  prospects on majoron dark matter}},\ }\href
  {https://doi.org/10.1007/JHEP07(2023)132} {\bibfield  {journal} {\bibinfo
  {journal} {JHEP}\ }\textbf {\bibinfo {volume} {07}},\ \bibinfo {pages}
  {132}},\ \Eprint {https://arxiv.org/abs/2304.04430} {arXiv:2304.04430
  [hep-ph]} \BibitemShut {NoStop}%
\bibitem [{\citenamefont {de~Giorgi}\ \emph {et~al.}(2024)\citenamefont
  {de~Giorgi}, \citenamefont {Merlo}, \citenamefont {Ponce~D\'\i{}az},\ and\
  \citenamefont {Rigolin}}]{deGiorgi:2023tvn}%
  \BibitemOpen
  \bibfield  {author} {\bibinfo {author} {\bibfnamefont {A.}~\bibnamefont
  {de~Giorgi}}, \bibinfo {author} {\bibfnamefont {L.}~\bibnamefont {Merlo}},
  \bibinfo {author} {\bibfnamefont {X.}~\bibnamefont {Ponce~D\'\i{}az}},\ and\
  \bibinfo {author} {\bibfnamefont {S.}~\bibnamefont {Rigolin}},\ }\bibfield
  {title} {\bibinfo {title} {{The minimal massive Majoron Seesaw Model}},\
  }\href {https://doi.org/10.1007/JHEP03(2024)094} {\bibfield  {journal}
  {\bibinfo  {journal} {JHEP}\ }\textbf {\bibinfo {volume} {03}},\ \bibinfo
  {pages} {094}},\ \Eprint {https://arxiv.org/abs/2312.13417} {arXiv:2312.13417
  [hep-ph]} \BibitemShut {NoStop}%
\bibitem [{\citenamefont {Heeck}\ and\ \citenamefont
  {Teresi}(2017)}]{Heeck:2017xbu}%
  \BibitemOpen
  \bibfield  {author} {\bibinfo {author} {\bibfnamefont {J.}~\bibnamefont
  {Heeck}}\ and\ \bibinfo {author} {\bibfnamefont {D.}~\bibnamefont {Teresi}},\
  }\bibfield  {title} {\bibinfo {title} {{Cold keV dark matter from decays and
  scatterings}},\ }\href {https://doi.org/10.1103/PhysRevD.96.035018}
  {\bibfield  {journal} {\bibinfo  {journal} {Phys. Rev. D}\ }\textbf {\bibinfo
  {volume} {96}},\ \bibinfo {pages} {035018} (\bibinfo {year} {2017})},\
  \Eprint {https://arxiv.org/abs/1706.09909} {arXiv:1706.09909 [hep-ph]}
  \BibitemShut {NoStop}%
\bibitem [{\citenamefont {Bloch}\ \emph {et~al.}(2021)\citenamefont {Bloch},
  \citenamefont {Caputo}, \citenamefont {Essig}, \citenamefont {Redigolo},
  \citenamefont {Sholapurkar},\ and\ \citenamefont {Volansky}}]{Bloch:2020uzh}%
  \BibitemOpen
  \bibfield  {author} {\bibinfo {author} {\bibfnamefont {I.~M.}\ \bibnamefont
  {Bloch}}, \bibinfo {author} {\bibfnamefont {A.}~\bibnamefont {Caputo}},
  \bibinfo {author} {\bibfnamefont {R.}~\bibnamefont {Essig}}, \bibinfo
  {author} {\bibfnamefont {D.}~\bibnamefont {Redigolo}}, \bibinfo {author}
  {\bibfnamefont {M.}~\bibnamefont {Sholapurkar}},\ and\ \bibinfo {author}
  {\bibfnamefont {T.}~\bibnamefont {Volansky}},\ }\bibfield  {title} {\bibinfo
  {title} {{Exploring new physics with O(keV) electron recoils in direct
  detection experiments}},\ }\href {https://doi.org/10.1007/JHEP01(2021)178}
  {\bibfield  {journal} {\bibinfo  {journal} {JHEP}\ }\textbf {\bibinfo
  {volume} {01}},\ \bibinfo {pages} {178}},\ \Eprint
  {https://arxiv.org/abs/2006.14521} {arXiv:2006.14521 [hep-ph]} \BibitemShut
  {NoStop}%
\bibitem [{\citenamefont {Lin}\ and\ \citenamefont
  {Yanagida}(2023)}]{Lin:2022niw}%
  \BibitemOpen
  \bibfield  {author} {\bibinfo {author} {\bibfnamefont {W.}~\bibnamefont
  {Lin}}\ and\ \bibinfo {author} {\bibfnamefont {T.~T.}\ \bibnamefont
  {Yanagida}},\ }\bibfield  {title} {\bibinfo {title} {{Consistency of the
  string inspired electroweak axion with cosmic birefringence}},\ }\href
  {https://doi.org/10.1103/PhysRevD.107.L021302} {\bibfield  {journal}
  {\bibinfo  {journal} {Phys. Rev. D}\ }\textbf {\bibinfo {volume} {107}},\
  \bibinfo {pages} {L021302} (\bibinfo {year} {2023})},\ \Eprint
  {https://arxiv.org/abs/2208.06843} {arXiv:2208.06843 [hep-ph]} \BibitemShut
  {NoStop}%
\bibitem [{\citenamefont {Langacker}\ \emph {et~al.}(1986)\citenamefont
  {Langacker}, \citenamefont {Peccei},\ and\ \citenamefont
  {Yanagida}}]{Langacker:1986rj}%
  \BibitemOpen
  \bibfield  {author} {\bibinfo {author} {\bibfnamefont {P.}~\bibnamefont
  {Langacker}}, \bibinfo {author} {\bibfnamefont {R.~D.}\ \bibnamefont
  {Peccei}},\ and\ \bibinfo {author} {\bibfnamefont {T.}~\bibnamefont
  {Yanagida}},\ }\bibfield  {title} {\bibinfo {title} {{Invisible Axions and
  Light Neutrinos: Are They Connected?}},\ }\href
  {https://doi.org/10.1142/S0217732386000683} {\bibfield  {journal} {\bibinfo
  {journal} {Mod. Phys. Lett. A}\ }\textbf {\bibinfo {volume} {1}},\ \bibinfo
  {pages} {541} (\bibinfo {year} {1986})}\BibitemShut {NoStop}%
\bibitem [{\citenamefont {Clarke}\ and\ \citenamefont
  {Volkas}(2016)}]{Clarke:2015bea}%
  \BibitemOpen
  \bibfield  {author} {\bibinfo {author} {\bibfnamefont {J.~D.}\ \bibnamefont
  {Clarke}}\ and\ \bibinfo {author} {\bibfnamefont {R.~R.}\ \bibnamefont
  {Volkas}},\ }\bibfield  {title} {\bibinfo {title} {{Technically natural
  nonsupersymmetric model of neutrino masses, baryogenesis, the strong CP
  problem, and dark matter}},\ }\href
  {https://doi.org/10.1103/PhysRevD.93.035001} {\bibfield  {journal} {\bibinfo
  {journal} {Phys. Rev. D}\ }\textbf {\bibinfo {volume} {93}},\ \bibinfo
  {pages} {035001} (\bibinfo {year} {2016})},\ \Eprint
  {https://arxiv.org/abs/1509.07243} {arXiv:1509.07243 [hep-ph]} \BibitemShut
  {NoStop}%
\bibitem [{\citenamefont {Sopov}\ and\ \citenamefont
  {Volkas}(2023)}]{Sopov:2022bog}%
  \BibitemOpen
  \bibfield  {author} {\bibinfo {author} {\bibfnamefont {A.~H.}\ \bibnamefont
  {Sopov}}\ and\ \bibinfo {author} {\bibfnamefont {R.~R.}\ \bibnamefont
  {Volkas}},\ }\bibfield  {title} {\bibinfo {title} {{VISH\ensuremath{\nu}:
  solving five Standard Model shortcomings with a Poincar\'e-protected
  electroweak scale}},\ }\href {https://doi.org/10.1016/j.dark.2023.101381}
  {\bibfield  {journal} {\bibinfo  {journal} {Phys. Dark Univ.}\ }\textbf
  {\bibinfo {volume} {42}},\ \bibinfo {pages} {101381} (\bibinfo {year}
  {2023})},\ \Eprint {https://arxiv.org/abs/2206.11598} {arXiv:2206.11598
  [hep-ph]} \BibitemShut {NoStop}%
\bibitem [{\citenamefont {Di~Luzio}\ \emph {et~al.}(2024)\citenamefont
  {Di~Luzio}, \citenamefont {Guerrera}, \citenamefont {Ponce~D\'\i{}az},\ and\
  \citenamefont {Rigolin}}]{DiLuzio:2024jip}%
  \BibitemOpen
  \bibfield  {author} {\bibinfo {author} {\bibfnamefont {L.}~\bibnamefont
  {Di~Luzio}}, \bibinfo {author} {\bibfnamefont {A.~W.~M.}\ \bibnamefont
  {Guerrera}}, \bibinfo {author} {\bibfnamefont {X.}~\bibnamefont
  {Ponce~D\'\i{}az}},\ and\ \bibinfo {author} {\bibfnamefont {S.}~\bibnamefont
  {Rigolin}},\ }\href@noop {} {\bibinfo {title} {{Axion-Like Particles in
  Radiative Quarkonia Decays}}} (\bibinfo {year} {2024}),\ \Eprint
  {https://arxiv.org/abs/2402.12454} {arXiv:2402.12454 [hep-ph]} \BibitemShut
  {NoStop}%
\bibitem [{\citenamefont {Clarke}\ \emph {et~al.}(2015)\citenamefont {Clarke},
  \citenamefont {Foot},\ and\ \citenamefont {Volkas}}]{Clarke:2015hta}%
  \BibitemOpen
  \bibfield  {author} {\bibinfo {author} {\bibfnamefont {J.~D.}\ \bibnamefont
  {Clarke}}, \bibinfo {author} {\bibfnamefont {R.}~\bibnamefont {Foot}},\ and\
  \bibinfo {author} {\bibfnamefont {R.~R.}\ \bibnamefont {Volkas}},\ }\bibfield
   {title} {\bibinfo {title} {{Natural leptogenesis and neutrino masses with
  two Higgs doublets}},\ }\href {https://doi.org/10.1103/PhysRevD.92.033006}
  {\bibfield  {journal} {\bibinfo  {journal} {Phys. Rev. D}\ }\textbf {\bibinfo
  {volume} {92}},\ \bibinfo {pages} {033006} (\bibinfo {year} {2015})},\
  \Eprint {https://arxiv.org/abs/1505.05744} {arXiv:1505.05744 [hep-ph]}
  \BibitemShut {NoStop}%
\bibitem [{\citenamefont {Blinov}\ \emph {et~al.}(2019)\citenamefont {Blinov},
  \citenamefont {Dolan}, \citenamefont {Draper},\ and\ \citenamefont
  {Kozaczuk}}]{Blinov:2019rhb}%
  \BibitemOpen
  \bibfield  {author} {\bibinfo {author} {\bibfnamefont {N.}~\bibnamefont
  {Blinov}}, \bibinfo {author} {\bibfnamefont {M.~J.}\ \bibnamefont {Dolan}},
  \bibinfo {author} {\bibfnamefont {P.}~\bibnamefont {Draper}},\ and\ \bibinfo
  {author} {\bibfnamefont {J.}~\bibnamefont {Kozaczuk}},\ }\bibfield  {title}
  {\bibinfo {title} {{Dark matter targets for axionlike particle searches}},\
  }\href {https://doi.org/10.1103/PhysRevD.100.015049} {\bibfield  {journal}
  {\bibinfo  {journal} {Phys. Rev. D}\ }\textbf {\bibinfo {volume} {100}},\
  \bibinfo {pages} {015049} (\bibinfo {year} {2019})},\ \Eprint
  {https://arxiv.org/abs/1905.06952} {arXiv:1905.06952 [hep-ph]} \BibitemShut
  {NoStop}%
\bibitem [{\citenamefont {Bardeen}\ \emph {et~al.}(1983)\citenamefont
  {Bardeen}, \citenamefont {Steinhardt},\ and\ \citenamefont
  {Turner}}]{Bardeen:1983qw}%
  \BibitemOpen
  \bibfield  {author} {\bibinfo {author} {\bibfnamefont {J.~M.}\ \bibnamefont
  {Bardeen}}, \bibinfo {author} {\bibfnamefont {P.~J.}\ \bibnamefont
  {Steinhardt}},\ and\ \bibinfo {author} {\bibfnamefont {M.~S.}\ \bibnamefont
  {Turner}},\ }\bibfield  {title} {\bibinfo {title} {{Spontaneous Creation of
  Almost Scale - Free Density Perturbations in an Inflationary Universe}},\
  }\href {https://doi.org/10.1103/PhysRevD.28.679} {\bibfield  {journal}
  {\bibinfo  {journal} {Phys. Rev. D}\ }\textbf {\bibinfo {volume} {28}},\
  \bibinfo {pages} {679} (\bibinfo {year} {1983})}\BibitemShut {NoStop}%
\bibitem [{\citenamefont {Seckel}\ and\ \citenamefont
  {Turner}(1985)}]{Seckel:1985tj}%
  \BibitemOpen
  \bibfield  {author} {\bibinfo {author} {\bibfnamefont {D.}~\bibnamefont
  {Seckel}}\ and\ \bibinfo {author} {\bibfnamefont {M.~S.}\ \bibnamefont
  {Turner}},\ }\bibfield  {title} {\bibinfo {title} {{Isothermal Density
  Perturbations in an Axion Dominated Inflationary Universe}},\ }\href
  {https://doi.org/10.1103/PhysRevD.32.3178} {\bibfield  {journal} {\bibinfo
  {journal} {Phys. Rev. D}\ }\textbf {\bibinfo {volume} {32}},\ \bibinfo
  {pages} {3178} (\bibinfo {year} {1985})}\BibitemShut {NoStop}%
\bibitem [{\citenamefont {O'Hare}(2020)}]{AxionLimits}%
  \BibitemOpen
  \bibfield  {author} {\bibinfo {author} {\bibfnamefont {C.}~\bibnamefont
  {O'Hare}},\ }\href {https://doi.org/10.5281/zenodo.3932430} {\bibinfo {title}
  {cajohare/axionlimits: Axionlimits}},\ \bibinfo {howpublished}
  {\url{https://cajohare.github.io/AxionLimits/}} (\bibinfo {year}
  {2020})\BibitemShut {NoStop}%
\bibitem [{\citenamefont {Du}\ \emph {et~al.}(2018)\citenamefont {Du} \emph
  {et~al.}}]{ADMX:2018gho}%
  \BibitemOpen
  \bibfield  {author} {\bibinfo {author} {\bibfnamefont {N.}~\bibnamefont {Du}}
  \emph {et~al.} (\bibinfo {collaboration} {ADMX}),\ }\bibfield  {title}
  {\bibinfo {title} {{A Search for Invisible Axion Dark Matter with the Axion
  Dark Matter Experiment}},\ }\href
  {https://doi.org/10.1103/PhysRevLett.120.151301} {\bibfield  {journal}
  {\bibinfo  {journal} {Phys. Rev. Lett.}\ }\textbf {\bibinfo {volume} {120}},\
  \bibinfo {pages} {151301} (\bibinfo {year} {2018})},\ \Eprint
  {https://arxiv.org/abs/1804.05750} {arXiv:1804.05750 [hep-ex]} \BibitemShut
  {NoStop}%
\bibitem [{\citenamefont {Braine}\ \emph {et~al.}(2020)\citenamefont {Braine}
  \emph {et~al.}}]{ADMX:2019uok}%
  \BibitemOpen
  \bibfield  {author} {\bibinfo {author} {\bibfnamefont {T.}~\bibnamefont
  {Braine}} \emph {et~al.} (\bibinfo {collaboration} {ADMX}),\ }\bibfield
  {title} {\bibinfo {title} {{Extended Search for the Invisible Axion with the
  Axion Dark Matter Experiment}},\ }\href
  {https://doi.org/10.1103/PhysRevLett.124.101303} {\bibfield  {journal}
  {\bibinfo  {journal} {Phys. Rev. Lett.}\ }\textbf {\bibinfo {volume} {124}},\
  \bibinfo {pages} {101303} (\bibinfo {year} {2020})},\ \Eprint
  {https://arxiv.org/abs/1910.08638} {arXiv:1910.08638 [hep-ex]} \BibitemShut
  {NoStop}%
\bibitem [{\citenamefont {Bartram}\ \emph {et~al.}(2021)\citenamefont {Bartram}
  \emph {et~al.}}]{ADMX:2021nhd}%
  \BibitemOpen
  \bibfield  {author} {\bibinfo {author} {\bibfnamefont {C.}~\bibnamefont
  {Bartram}} \emph {et~al.} (\bibinfo {collaboration} {ADMX}),\ }\bibfield
  {title} {\bibinfo {title} {{Search for Invisible Axion Dark Matter in the
  3.3\textendash{}4.2\,\,\ensuremath{\mu}eV Mass Range}},\ }\href
  {https://doi.org/10.1103/PhysRevLett.127.261803} {\bibfield  {journal}
  {\bibinfo  {journal} {Phys. Rev. Lett.}\ }\textbf {\bibinfo {volume} {127}},\
  \bibinfo {pages} {261803} (\bibinfo {year} {2021})},\ \Eprint
  {https://arxiv.org/abs/2110.06096} {arXiv:2110.06096 [hep-ex]} \BibitemShut
  {NoStop}%
\bibitem [{\citenamefont {Ahyoune}\ \emph {et~al.}(2023)\citenamefont {Ahyoune}
  \emph {et~al.}}]{Ahyoune:2023gfw}%
  \BibitemOpen
  \bibfield  {author} {\bibinfo {author} {\bibfnamefont {S.}~\bibnamefont
  {Ahyoune}} \emph {et~al.},\ }\bibfield  {title} {\bibinfo {title} {{A
  Proposal for a Low-Frequency Axion Search in the 1\textendash{}2
  \ensuremath{\mu}eV Range and Below with the BabyIAXO Magnet}},\ }\href
  {https://doi.org/10.1002/andp.202300326} {\bibfield  {journal} {\bibinfo
  {journal} {Annalen Phys.}\ }\textbf {\bibinfo {volume} {535}},\ \bibinfo
  {pages} {2300326} (\bibinfo {year} {2023})},\ \Eprint
  {https://arxiv.org/abs/2306.17243} {arXiv:2306.17243 [physics.ins-det]}
  \BibitemShut {NoStop}%
\bibitem [{\citenamefont {Alesini}\ \emph {et~al.}(2023)\citenamefont {Alesini}
  \emph {et~al.}}]{Alesini:2023qed}%
  \BibitemOpen
  \bibfield  {author} {\bibinfo {author} {\bibfnamefont {D.}~\bibnamefont
  {Alesini}} \emph {et~al.},\ }\bibfield  {title} {\bibinfo {title} {{The
  future search for low-frequency axions and new physics with the FLASH
  resonant cavity experiment at Frascati National Laboratories}},\ }\href
  {https://doi.org/10.1016/j.dark.2023.101370} {\bibfield  {journal} {\bibinfo
  {journal} {Phys. Dark Univ.}\ }\textbf {\bibinfo {volume} {42}},\ \bibinfo
  {pages} {101370} (\bibinfo {year} {2023})},\ \Eprint
  {https://arxiv.org/abs/2309.00351} {arXiv:2309.00351 [physics.ins-det]}
  \BibitemShut {NoStop}%
\bibitem [{\citenamefont {Berlin}\ \emph {et~al.}(2021)\citenamefont {Berlin},
  \citenamefont {D'Agnolo}, \citenamefont {Ellis},\ and\ \citenamefont
  {Zhou}}]{Berlin:2020vrk}%
  \BibitemOpen
  \bibfield  {author} {\bibinfo {author} {\bibfnamefont {A.}~\bibnamefont
  {Berlin}}, \bibinfo {author} {\bibfnamefont {R.~T.}\ \bibnamefont
  {D'Agnolo}}, \bibinfo {author} {\bibfnamefont {S.~A.~R.}\ \bibnamefont
  {Ellis}},\ and\ \bibinfo {author} {\bibfnamefont {K.}~\bibnamefont {Zhou}},\
  }\bibfield  {title} {\bibinfo {title} {{Heterodyne broadband detection of
  axion dark matter}},\ }\href {https://doi.org/10.1103/PhysRevD.104.L111701}
  {\bibfield  {journal} {\bibinfo  {journal} {Phys. Rev. D}\ }\textbf {\bibinfo
  {volume} {104}},\ \bibinfo {pages} {L111701} (\bibinfo {year} {2021})},\
  \Eprint {https://arxiv.org/abs/2007.15656} {arXiv:2007.15656 [hep-ph]}
  \BibitemShut {NoStop}%
\bibitem [{\citenamefont {Brouwer}\ \emph {et~al.}(2022)\citenamefont {Brouwer}
  \emph {et~al.}}]{DMRadio:2022pkf}%
  \BibitemOpen
  \bibfield  {author} {\bibinfo {author} {\bibfnamefont {L.}~\bibnamefont
  {Brouwer}} \emph {et~al.} (\bibinfo {collaboration} {DMRadio}),\ }\bibfield
  {title} {\bibinfo {title} {{Projected sensitivity of DMRadio-m3: A search for
  the QCD axion below 1\,\,\ensuremath{\mu}eV}},\ }\href
  {https://doi.org/10.1103/PhysRevD.106.103008} {\bibfield  {journal} {\bibinfo
   {journal} {Phys. Rev. D}\ }\textbf {\bibinfo {volume} {106}},\ \bibinfo
  {pages} {103008} (\bibinfo {year} {2022})},\ \Eprint
  {https://arxiv.org/abs/2204.13781} {arXiv:2204.13781 [hep-ex]} \BibitemShut
  {NoStop}%
\bibitem [{\citenamefont {Nomura}\ \emph {et~al.}(2000)\citenamefont {Nomura},
  \citenamefont {Watari},\ and\ \citenamefont {Yanagida}}]{Nomura:2000yk}%
  \BibitemOpen
  \bibfield  {author} {\bibinfo {author} {\bibfnamefont {Y.}~\bibnamefont
  {Nomura}}, \bibinfo {author} {\bibfnamefont {T.}~\bibnamefont {Watari}},\
  and\ \bibinfo {author} {\bibfnamefont {T.}~\bibnamefont {Yanagida}},\
  }\bibfield  {title} {\bibinfo {title} {{Quintessence axion potential induced
  by electroweak instanton effects}},\ }\href
  {https://doi.org/10.1016/S0370-2693(00)00605-5} {\bibfield  {journal}
  {\bibinfo  {journal} {Phys. Lett. B}\ }\textbf {\bibinfo {volume} {484}},\
  \bibinfo {pages} {103} (\bibinfo {year} {2000})},\ \Eprint
  {https://arxiv.org/abs/hep-ph/0004182} {arXiv:hep-ph/0004182} \BibitemShut
  {NoStop}%
\bibitem [{\citenamefont {Choi}\ \emph {et~al.}(2021)\citenamefont {Choi},
  \citenamefont {Lin}, \citenamefont {Visinelli},\ and\ \citenamefont
  {Yanagida}}]{Choi:2021aze}%
  \BibitemOpen
  \bibfield  {author} {\bibinfo {author} {\bibfnamefont {G.}~\bibnamefont
  {Choi}}, \bibinfo {author} {\bibfnamefont {W.}~\bibnamefont {Lin}}, \bibinfo
  {author} {\bibfnamefont {L.}~\bibnamefont {Visinelli}},\ and\ \bibinfo
  {author} {\bibfnamefont {T.~T.}\ \bibnamefont {Yanagida}},\ }\bibfield
  {title} {\bibinfo {title} {{Cosmic birefringence and electroweak axion dark
  energy}},\ }\href {https://doi.org/10.1103/PhysRevD.104.L101302} {\bibfield
  {journal} {\bibinfo  {journal} {Phys. Rev. D}\ }\textbf {\bibinfo {volume}
  {104}},\ \bibinfo {pages} {L101302} (\bibinfo {year} {2021})},\ \Eprint
  {https://arxiv.org/abs/2106.12602} {arXiv:2106.12602 [hep-ph]} \BibitemShut
  {NoStop}%
\bibitem [{\citenamefont {Minami}\ and\ \citenamefont
  {Komatsu}(2020)}]{Minami:2020odp}%
  \BibitemOpen
  \bibfield  {author} {\bibinfo {author} {\bibfnamefont {Y.}~\bibnamefont
  {Minami}}\ and\ \bibinfo {author} {\bibfnamefont {E.}~\bibnamefont
  {Komatsu}},\ }\bibfield  {title} {\bibinfo {title} {{New Extraction of the
  Cosmic Birefringence from the Planck 2018 Polarization Data}},\ }\href
  {https://doi.org/10.1103/PhysRevLett.125.221301} {\bibfield  {journal}
  {\bibinfo  {journal} {Phys. Rev. Lett.}\ }\textbf {\bibinfo {volume} {125}},\
  \bibinfo {pages} {221301} (\bibinfo {year} {2020})},\ \Eprint
  {https://arxiv.org/abs/2011.11254} {arXiv:2011.11254 [astro-ph.CO]}
  \BibitemShut {NoStop}%
\bibitem [{\citenamefont {Diego-Palazuelos}\ \emph {et~al.}(2022)\citenamefont
  {Diego-Palazuelos} \emph {et~al.}}]{Diego-Palazuelos:2022dsq}%
  \BibitemOpen
  \bibfield  {author} {\bibinfo {author} {\bibfnamefont {P.}~\bibnamefont
  {Diego-Palazuelos}} \emph {et~al.},\ }\bibfield  {title} {\bibinfo {title}
  {{Cosmic Birefringence from the Planck Data Release 4}},\ }\href
  {https://doi.org/10.1103/PhysRevLett.128.091302} {\bibfield  {journal}
  {\bibinfo  {journal} {Phys. Rev. Lett.}\ }\textbf {\bibinfo {volume} {128}},\
  \bibinfo {pages} {091302} (\bibinfo {year} {2022})},\ \Eprint
  {https://arxiv.org/abs/2201.07682} {arXiv:2201.07682 [astro-ph.CO]}
  \BibitemShut {NoStop}%
\bibitem [{\citenamefont {Frampton}\ \emph {et~al.}(2002)\citenamefont
  {Frampton}, \citenamefont {Glashow},\ and\ \citenamefont
  {Yanagida}}]{Frampton:2002qc}%
  \BibitemOpen
  \bibfield  {author} {\bibinfo {author} {\bibfnamefont {P.~H.}\ \bibnamefont
  {Frampton}}, \bibinfo {author} {\bibfnamefont {S.~L.}\ \bibnamefont
  {Glashow}},\ and\ \bibinfo {author} {\bibfnamefont {T.}~\bibnamefont
  {Yanagida}},\ }\bibfield  {title} {\bibinfo {title} {{Cosmological sign of
  neutrino CP violation}},\ }\href
  {https://doi.org/10.1016/S0370-2693(02)02853-8} {\bibfield  {journal}
  {\bibinfo  {journal} {Phys. Lett. B}\ }\textbf {\bibinfo {volume} {548}},\
  \bibinfo {pages} {119} (\bibinfo {year} {2002})},\ \Eprint
  {https://arxiv.org/abs/hep-ph/0208157} {arXiv:hep-ph/0208157} \BibitemShut
  {NoStop}%
\bibitem [{\citenamefont {Di~Luzio}\ \emph {et~al.}(2023)\citenamefont
  {Di~Luzio}, \citenamefont {Guerrera}, \citenamefont {D\'\i{}az},\ and\
  \citenamefont {Rigolin}}]{DiLuzio:2023ndz}%
  \BibitemOpen
  \bibfield  {author} {\bibinfo {author} {\bibfnamefont {L.}~\bibnamefont
  {Di~Luzio}}, \bibinfo {author} {\bibfnamefont {A.~W.~M.}\ \bibnamefont
  {Guerrera}}, \bibinfo {author} {\bibfnamefont {X.~P.}\ \bibnamefont
  {D\'\i{}az}},\ and\ \bibinfo {author} {\bibfnamefont {S.}~\bibnamefont
  {Rigolin}},\ }\bibfield  {title} {\bibinfo {title} {{On the IR/UV flavour
  connection in non-universal axion models}},\ }\href
  {https://doi.org/10.1007/JHEP06(2023)046} {\bibfield  {journal} {\bibinfo
  {journal} {JHEP}\ }\textbf {\bibinfo {volume} {06}},\ \bibinfo {pages}
  {046}},\ \Eprint {https://arxiv.org/abs/2304.04643} {arXiv:2304.04643
  [hep-ph]} \BibitemShut {NoStop}%
\end{thebibliography}%

\end{document}